\documentclass[twocolumn, fleqn]{aastex701}

\usepackage{subcaption} 
\usepackage{enumitem} 

\newcommand{\eq}[1]{eq.~(\ref{eq:#1})}

\newcommand{\se}[1]{Section \ref{sec:#1}}

\newcommand{\Fig}[1]{Fig.~\ref{fig:#1}}

\newcommand{\Tab}[1]{Table~\ref{tab:#1}}
\newcommand{\be}{\begin{equation}}
\newcommand{\ee}{\end{equation}}
\newcommand{\bad}{\begin{equation} \begin{aligned}}
\newcommand{\ead}{\end{aligned} \end{equation}}

\newcommand{\Msun}{M_\odot}

\newcommand{\g}{\,{\rm g}}

\newcommand{\Mv}{M_{\rm vir}}

\newcommand{\Rv}{R_{\rm vir}}

\begin{document}

\title{Universal Dark-matter Density Profiles of Cosmic Filaments}

\author[orcid=0009-0000-4754-5364,gname=Peng, sname=Xu]{Peng Xu}
\affiliation{Kavli Institute for Astronomy and Astrophysics, Peking University, Beijing 100871, People's Republic of China}
\affiliation{Zhili College, Tsinghua University, Beijing 100084, People's Republic of China}
\email[show]{\href{mailto:pengxu.thu@gmail.com}{pengxu.thu@gmail.com}}

\author[orcid=0000-0001-6115-0633,gname=Fangzhou, sname=Jiang]{Fangzhou Jiang} 
\altaffiliation{Corresponding author}
\affiliation{Kavli Institute for Astronomy and Astrophysics, Peking University, Beijing 100871, People's Republic of China}
\email[show]{\href{mailto:fangzhou.jiang@pku.edu.cn}{fangzhou.jiang@pku.edu.cn}}

\author[orcid=0000-0002-0072-0281, gname=Farhanul, sname=Hasan]{Farhanul Hasan}
\affiliation{Space Telescope Science Institute, 3700 San Martin Drive, Baltimore, MD 21218, USA}
\affiliation{Department of Astronomy, New Mexico State University, Las Cruces, NM 88003, USA}
\email{fhasan@stsci.edu}

\author[orcid=0000-0002-0072-0281, gname=Joanna, sname=Woo]{Joanna Woo}
\affiliation{Department of Physics, Simon Fraser University, 8888 University Drive, Burnaby, BC, V5A 1S6, Canada}
\email{j_woo@sfu.ca}

\author[orcid=0009-0005-7451-0614, gname=Douglas, sname=Hellinger]{Douglas Hellinger}
\affiliation{Santa Cruz Institute for Particle Physics, University of California, Santa Cruz, CA 95064, USA}
\email{dhelling@ucsc.edu}

\author[orcid=0000-0001-5091-5098,gname=Joel, sname=Primack]{Joel R. Primack}
\affiliation{Department of Physics, University of California, Santa Cruz, CA 95064, USA}
\email{joel@ucsc.edu}

\author[orcid=0000-0003-3385-6799, gname=David, sname=Koo]{David C. Koo}
\affiliation{UCO/Lick Observatory, Department of Astronomy and Astrophysics, University of California, Santa Cruz, CA 95064, USA}
\email{koo@ucolick.org}

\begin{abstract}
We present a comprehensive analysis of the radial dark-matter (DM) density profiles of cosmic filaments in the hydrodynamical simulation TNG50. 
The cosmic web is extracted from high-resolution density grids at redshifts $z =$ 0, 0.5, 1, 2 and 3 using the DisPerSE algorithm. 
We show that the filament spine locations returned directly by DisPerSE do not accurately reflect the true density ridges. 
To address this issue, we introduce a ``shrinking-cylinder" re-centering algorithm, which significantly increases the inferred central densities and restores the inner power-law behavior of the profiles.
When the radial coordinate is scaled by the virial radii of the terminal nodes, the filament density profiles exhibit a nearly universal form that depends only weakly on redshift, node mass, and filament length. This result suggests that cosmic filaments, much like dark-matter halos, obey a form of structural self-similarity once an appropriate characteristic scale is introduced.
By repeating the measurement using only smoothly distributed, unbound DM particles, we find that the apparent central cusp of the full profile is primarily produced by low-mass halos embedded along the filament spines, while the smooth component develops a flat core within $R/R_{\rm vir}\lesssim0.1$. 
The redshift evolution of this smooth component further suggests a transition from predominantly smooth filamentary accretion at high redshift to increasingly clumpy accretion at late times.
Finally, we show that the universal filament profile is accurately described by a generalized triple-power-law model.
\end{abstract}



\section{Introduction}\label{sec:intro}

On megaparsec scales, the matter and galaxy distributions in the Universe are not uniform, but instead form an intricate, multi-scale interconnected network known as the \textit{cosmic web} \citep{Bond96}. 
This network is arranged into a salient pattern of dense clusters (nodes), elongated filaments, and extended walls surrounding near-empty voids. 
Filaments facilitate the accretion and transport of mass, intergalactic gas, and galaxies toward high-density cluster regions \citep{vanHaarlem93, Knebe04}. 
Since galaxies are commonly found within or near these structures, filaments may provide an important context for studying the environmental dependence of galaxy properties.
However, filaments remain far less well characterized in both cosmological simulations and observations than cosmic nodes (halos), largely because they lack a unique physical definition and boundary, as well as robust observational tracers.

Since structure formation in the standard paradigm is hierarchical, the cosmic web spans a wide range of scales.  
By analogy with dark-matter (DM) halos, which exhibit nearly universal density-profile shapes across epochs and mass scales \citep{NFW97} while still retaining systematic variations in their internal structures \citep{Allgood06, Dutton14}, it is natural to speculate that filaments of different scales (for example, those connected to halos of different masses) may similarly exhibit self-similarity while differing subtly in their structural properties.

Existing quantitative studies of cosmic filaments cover a wide range of topics, yet most analyses treat filaments as a single population, overlooking their multi-scale nature, and therefore characterize them only on the largest scales \citep{Gheller16, Zhu21, Galarraga-Espinosa24, Bahe25, Yang25}. 
It is therefore essential to explicitly define the relevant mass, spatial, and temporal scales of interest, and to characterize filamentary structures as a function of these scales.

A number of filament (cosmic web) finders have been developed based on a variety of mathematical and physical concepts \citep[see][for a detailed review]{Libeskind18}.
Among them, the topological DisPerSE algorithm \citep{Sousbie11a, Sousbie11b} is one of the most widely used methods.
However, the algorithm involves user-defined choices whose impact on the resulting filament catalogs remains insufficiently documented in the literature and may lead to different interpretations or systematic biases.
For example, DisPerSE takes a density field as input for web extraction, provided either on a regular grid or as a discrete field reconstructed from tracers using its built-in DTFE interpolation. 
In practice, the latter is more commonly adopted, where galaxies above a given stellar-mass threshold serve as tracers. 
Recent works have shown that DTFE performs poorly in recovering the underlying density field \citep{Hasan24}, thereby introducing significant biases in the extracted cosmic web. 
In addition, the spine positions returned by DisPerSE can be offset because of the finite resolution of the input density grid, which may further bias the inferred density profiles of filaments. 
Establishing a consistent and standardized pipeline for filament extraction should be a foundational step for any study based on the DisPerSE algorithm.

Motivated by these goals, in this work, we develop a robust and accurate method to extract the DM cosmic web from cosmological simulations with the DisPerSE algorithm, and characterize the radial density profiles of cosmic filaments. 
In \se{method}, we illustrate the effects of inaccurate filament-spine centering in DisPerSE and introduce a re-centering algorithm to correct this effect. 
In \se{profile}, we quantify the radial DM density profiles around cosmic filaments and examine their dependence on physical properties such as node mass and filament length across multiple cosmic epochs. 
We also present empirical models that can describe the profiles of filaments.
We discuss the implications of our results in \se{discussion} and conclude the paper in \se{conclusion}.

\section{Data and Methods}\label{sec:method}

\subsection{TNG Simluations}
We analyze the gravo-magnetohydrodynamical simulation IllustrisTNG\footnote{\url{https://www.tng-project.org}} \citep{Nelson18, Nelson19, Pillepich18}, which was carried out with the moving-mesh code Arepo \citep{Springel10} to follow the evolution of DM, gas, stars and blackholes from redshift $z$ = 127 to $z$ = 0. 
The adopted cosmological constants are consistent with the \textit{Planck} 2015 \citep{Planck16} results: $\Omega_{\Lambda, 0}$ = 0.6911, $\Omega_{m, 0}$ = 0.3089, $\Omega_{b, 0}$ = 0.0486, $\sigma_8$ = 0.8159, $n_s$ = 0.9667 and $h$ = 0.6774. 
We focus on the highest-resolution box, TNG50-1, at redshifts $z$ = 0, 0.5, 1, 2, and 3, in order to accurately characterize filaments down to small scales and out to high redshift. 
The simulation volume has a side length of 51.7 comoving Mpc and contains $2160^3$ DM particles of mass $4.5\times10^5 \Msun$. 
Our analysis of cosmic filaments also makes use of the DM halos connected by the filaments. 
We therefore use the publicly available group catalogs constructed with a friends-of-friends (FoF) algorithm, together with the subhalo catalogs generated using the SUBFIND algorithm \citep{Springel01, Dolag09}.

Starting from the coordinates of individual DM particles, we reconstruct the DM overdensity field, $\rho/\rho_0 = 1 + \delta$, where $\rho_0$ denotes the cosmic mean density, on a regular $512^3$ grid by assigning the mass of a particle uniformly to its hosting cell. 
The resulting density grid is then smoothed to suppress sampling noise. 
Following \cite{Cautun13}, we apply Gaussian smoothing to $\log(1+\delta)$ with $\sigma = 0.2$ cMpc, corresponding to an effective smoothing scale of $2\sqrt{2\ln 2}\,\sigma \approx 0.47$ cMpc (the Full Width at Half Maximum of the Gaussian kernel). 
Although applying a Gaussian filter to the logarithmic density field does not conserve mass, it has the advantage of enhancing lower-density filamentary structures relative to dense cosmic nodes, compared with smoothing the linear density field.

\subsection{An improved workflow of filament extraction using the DisPerSE Algorithm} \label{sec:DisPerSE}

\begin{figure}
    \centering
    \includegraphics[width=1\linewidth]{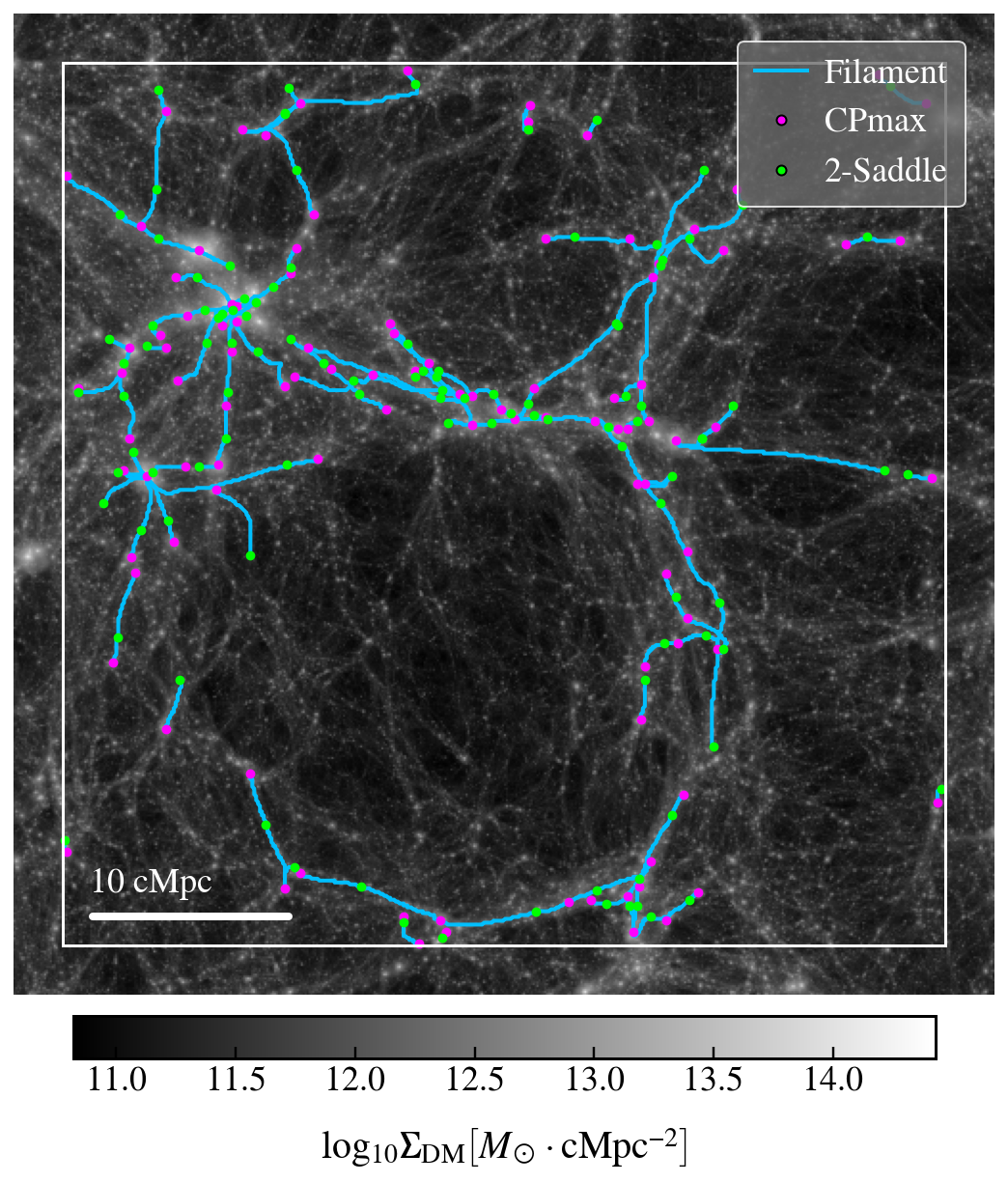}
    \caption{
    Illustration of the cosmic filament network identified by DisPerSE at $z$ = 1 in TNG50-1. 
    Blue lines show the filaments in a 20 cMpc-thick slice, while purple and green points mark maxima and saddle points, respectively. 
    The background image shows the DM surface density projected over the same slice. 
    Our analysis is restricted to the area enclosed by the white box, excluding the outer $\sim 5\%$ of the simulation volume to avoid boundary effects.}
    \label{fig:illustration}
\end{figure}

We extract filaments using the Discrete Persistent Structure Extractor \citep[DisPerSE\footnote{\url{https://www2.iap.fr/users/sousbie/web/html/indexd41d.html}},][]{Sousbie11a, Sousbie11b}. 
Its primary purpose is to identify persistent topological features, such as peaks, voids, walls, and filamentary structures, using discrete Morse theory \citep{Milnor16, Jost08}. 
By computing the gradient field of the input densities, the main DisPerSE routine, \textit{mse}, identifies critical points where the gradient vanishes. 
A critical index (ranging from 0 to 3) is attached to each critical point according to the eigenvalues of the Hessian matrix. 
Filaments are then traced by integral lines originating from maxima (CPmax, critical points with index 3) and terminating at saddle points (critical points with index 2).
Each filament consists of a sequence of connected segments.
The robustness of these structures is gauged by a topological quantity named \textit{persistence}, defined as the density ratio between paired critical points. 
In practice, a persistence threshold is applied to suppress noise during the structure-identification process.

We adopt different persistence thresholds ({\tt{cut}}, as opposed to the commonly used {\tt{nsig}} for DTFE density field estimated by tracers) for the gridded density inputs at different redshifts. 
The threshold decreases toward higher redshift, reflecting the fact that the density contrast grows with cosmic time. 
This parameter strongly affects the extracted filament network, with higher-persistence structures generally being more significant and robust than lower-persistence ones.
However, because our primary goal is to probe the multiscale nature of the cosmic web and the properties of filaments across different physical scales, we adopt relatively low thresholds at high redshift in order to better recover small-scale structures.
Importantly, these threshold choices are not arbitrary. Instead, they are calibrated to ensure that the most massive halos at each redshift are consistently associated with filament CPmax points (see below and Appendix \ref{AppendixA}).

Since the program runs on a regular grid, the filaments are naturally jagged and require smoothing. 
We choose $N_{\rm{smooth}}=15$ in the \textit{skelconv} program of DisPerSE; that is, the filament sampling points are replaced by the average of their neighboring points for 15 iterations. 
This smoothing operation is inherited from the default setting of DisPerSE rather than an additional processing step introduced in this work.
The resulting filament statistics are insensitive to the exact choice of $N_{\rm{smooth}}$. 
For a more detailed discussion of this number, see \cite{Bahe25}.

To avoid spurious maxima far from any density peak and extremely short or spurious filaments, we apply several post-processing procedures to refine the raw DisPerSE output. 
Following \cite{Galarraga-Espinosa20}, we first conservatively exclude the outer $\sim$5\% of the TNG50 box on each side to avoid boundary effects of topological calculations. 
We further remove filaments shorter than 0.4 cMpc, approximately twice the effective resolution limit of the smoothed density field. 
Each maximum of the remaining filaments is then associated with an FoF group more massive than $10^8\,\Msun$ if it is within that halo's virial radius.
Unmatched maxima and corresponding filaments are discarded.
This mass threshold is chosen to ensure that the halo is resolved with more than 100 DM particles.

\Fig{illustration} illustrates the outcome of the filament identification procedure described above, showing a slice of the filament network at $z = 1$. 
As can be seen, the identified filament spines trace the filamentary structures of the DM density field well. 
Because our focus is on precise characterization of filament structures, we prioritize sample cleanliness over completeness.

\subsection{Filament spine location refinement}\label{sec:recentering}

\begin{figure*}
    \centering
    \includegraphics[width=0.55\linewidth]{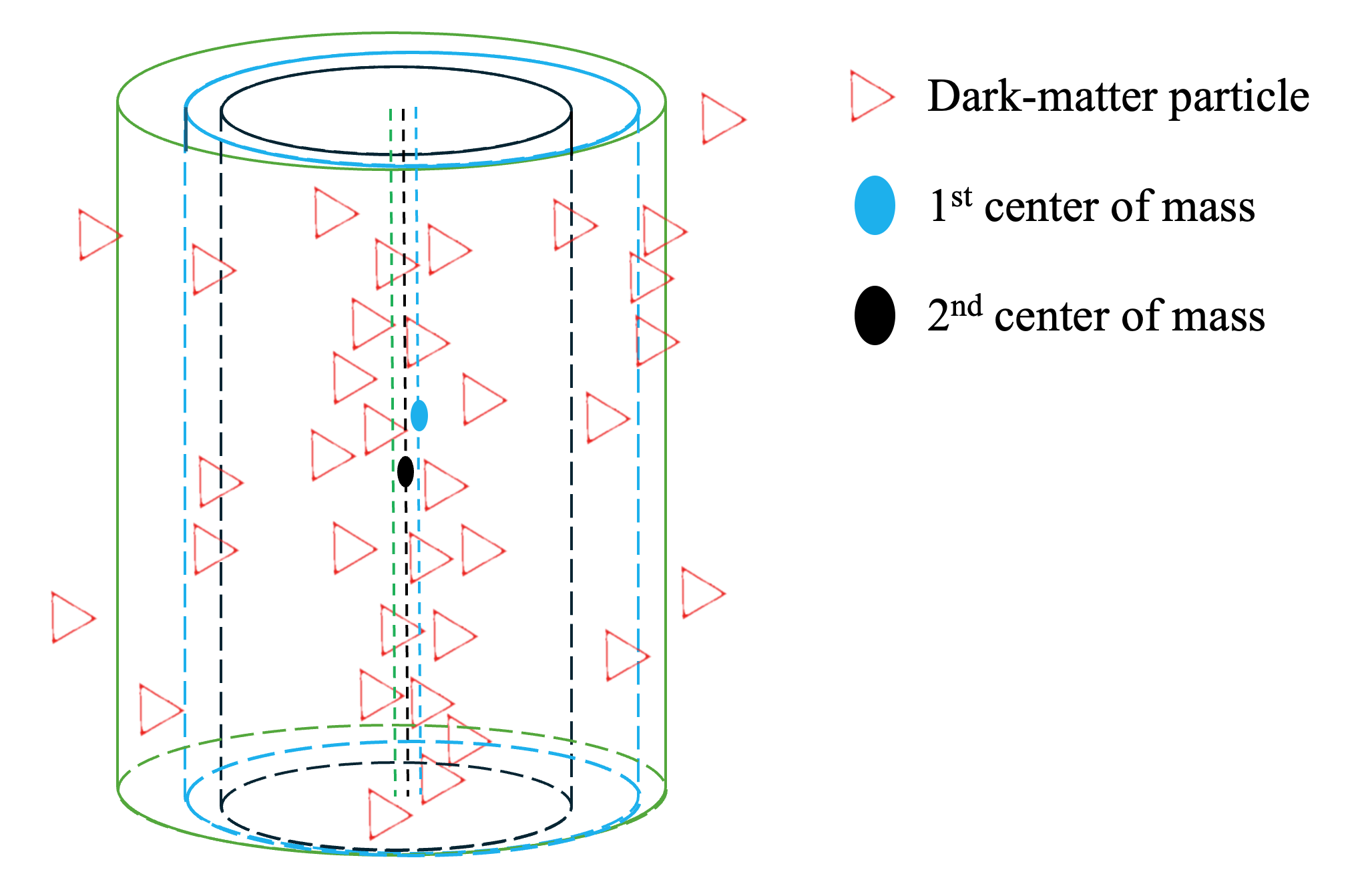}
    \includegraphics[width=0.4\linewidth]{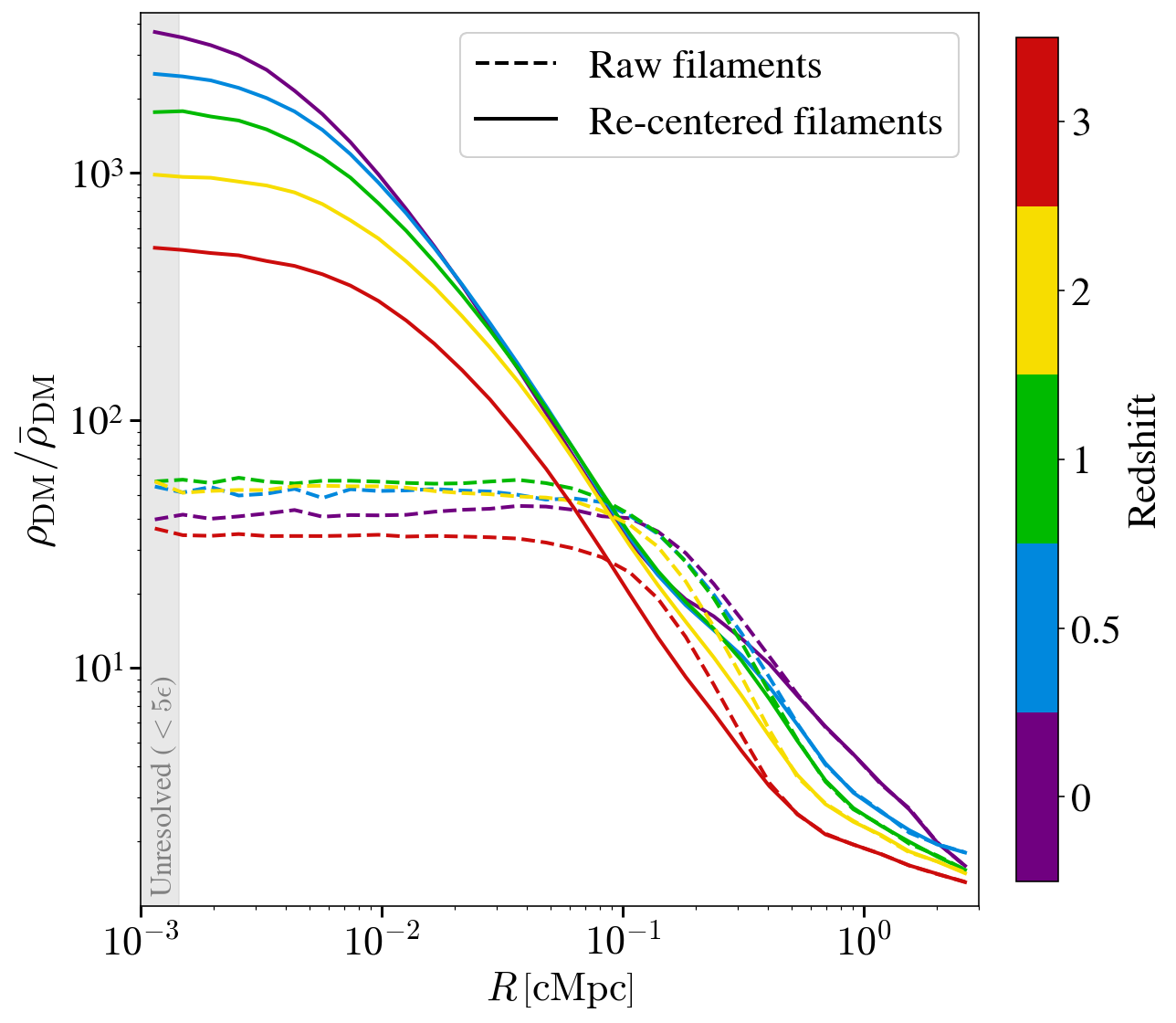}
    \caption{
    {\textit{Left}}: Illustration of the method used to refine the spine location of a filament segment. 
    The sequence of ``shrinking cylinder'' operations is indicated by the green, blue, and black cylinders. 
    At the $(k+1)$th iteration, the spine is repositioned to pass through the center of mass of particles enclosed within the cylindrical region of radius $R_k$ from the $k$th iteration, after which the radius is reduced by 10\%. 
    The iteration terminates when the number of enclosed particles falls below 100 or when the spine displacement between successive steps becomes smaller than 1 ckpc. 
    The allowed translations are restricted to directions perpendicular to the spine, i.e., rotations are not permitted.
    \quad
    {\textit{Right}}: Radial DM density profiles around the raw (dashed) and re-centered (solid) filaments at different redshifts. 
    All profiles are length-weighted means over the corresponding filament samples and are normalized by the cosmic mean density. 
    The radial range is limited to $R<3$ cMpc in order to minimize the impact of profiles extending beyond the simulation boundaries.
    The gray shaded region marks unresolved scales, defined as distances to the filament spine smaller than 5 times the DM softening length of TNG50-1.}
    \label{fig:recentering}
\end{figure*}

Because the topological calculations are based on input gridded density fields with finite spatial resolution, the spine positions returned by DisPerSE do not necessarily coincide with the true location of maximum density. 
To correct for this effect, the spine of each individual segment is refined using an iterative procedure in which the center of mass of particles within a shrinking cylinder is computed recursively until a convergence criterion is met. 
The re-centering process is illustrated in the left panel of \Fig{recentering}. 
To reduce the computational cost, we first randomly select 2\% of the DM particles after loading the particle data. This fixed subsample is then used throughout for both the re-centering and density profile calculations.
We verify that the resulting density profiles are insensitive to the subsampling fraction by comparing the results obtained using 2\% and 10\% of the DM particles, which show statistically indistinguishable results.
At each iteration, the cylinder axis is reset to pass through the barycenter obtained in the previous step, and the cylinder radius is reduced by 10\%. 
The initial radius is set to 0.2 cMpc, and the iteration terminates when the number of enclosed particles falls below 100 or when the displacement between successive iterations falls below 1 ckpc. 
This ``shrinking cylinder'' algorithm is inspired by the method widely used for centering galaxies or DM halos by calculating the center of mass within shrinking spheres \citep{Power03}. 
A similar approach has been adopted in previous studies of high-$z$ filament cross-sections, where an iterative procedure is applied to zoom-in slices of filaments \citep{Lu23}.
We find that this refinement leads to typical spine displacements of up to 20 ckpc.

\subsection{Definition of DM density profile around cosmic filaments}
\label{sec:DensityDefinition}

We compute the radial DM density profile around filaments as the volume density in concentric cylindrical shells. 
The profile around a segment $i$ is estimated from a set of concentric cylindrical shells around its axis, according to
\begin{equation}
    \rho_i^k = \frac{M_{\mathrm{DM},k}}{\pi L_i\left(R_k^{2}-R_{k-1}^{2}\right)} \, ,
\end{equation}
where $L_i$ is the length of segment $i$, and $R_k - R_{k-1}$ and $M_{\mathrm{DM},k}$ denote the thickness and enclosed DM mass of the $k$th  shell, respectively. 
A filament consists of a sequence of connected segments, originating from a CPmax within a node (halo) and ending at a saddle point.
The profile of an individual filament is then computed as the length-weighted mean of the profiles of its constituent segments. 
Finally, we take the length-weighted average over the full filament population to obtain the mean filament profile.
To avoid contamination from the connecting nodes, we exclude contributions from the node outskirts by discarding all regions within $\Rv$ of the terminal halos.

\section{Dark-matter density profiles of cosmic filaments}\label{sec:profile}

\subsection{Importance of filament spine re-centering}
\label{sec:RecenteringEffect}

To demonstrate the importance of spine refinement, we show in the right panel of \Fig{recentering} the filament DM density profiles derived using our re-centering procedure, and compare them with those obtained using the raw spine locations returned directly by DisPerSE. 
The densities are normalized to the comoving cosmic mean DM density, $\bar{\rho}_{\rm DM}$.
Relative to the unrefined results, the refined spine locations yield significantly higher central densities at all redshifts, by as much as 1-1.5 dex.
In addition, the refined profiles exhibit a nearly universal power-law behavior down to the innermost $\sim 0.01$ cMpc, substantially extending the radial range over which the power law holds. 
The radius where the profiles flatten ($\sim 0.01$ cMpc) is also consistent with results from \cite{Ramsoy21} that explicitly resolve and characterize small-scale filamentary structures.

These results demonstrate that using the spine locations output directly by DisPerSE can introduce significant systematic biases in the density profiles. 
This inaccuracy can be naturally attributed to the finite resolution of the input density grids.
Because DisPerSE constructs the Morse-Smale complex from the density values assigned to grid cells, it cannot recover the information that has been smoothed out in the density field. 
The re-centered profiles begin to deviate from the original ones at around 0.4-0.5 cMpc, exactly corresponding to the resolution limit of our density grid. 
All filament profiles presented in the remainder of this work therefore use the re-centered spines.

\subsection{Universality of density profiles normalized by node radius}
\label{sec:universality}

In defining and quantitatively characterizing cosmic structures, it is essential to first identify the physically relevant mass scale.
Because filaments are not virialized structures themselves, a natural choice for their characteristic mass scale is that of the neighboring virialized objects, namely, the dark-matter halos at their endpoints. A similar scaling approach has also been adopted in several previous studies to investigate the properties and dynamics of gaseous cosmic filaments \citep{Birnboim16, Mandelker18, Mandelker20, Lu23, Aung24}.

\begin{figure*}
    \centering

    \begin{subfigure}{0.68\textwidth}
        \centering
        \includegraphics[width=\linewidth]{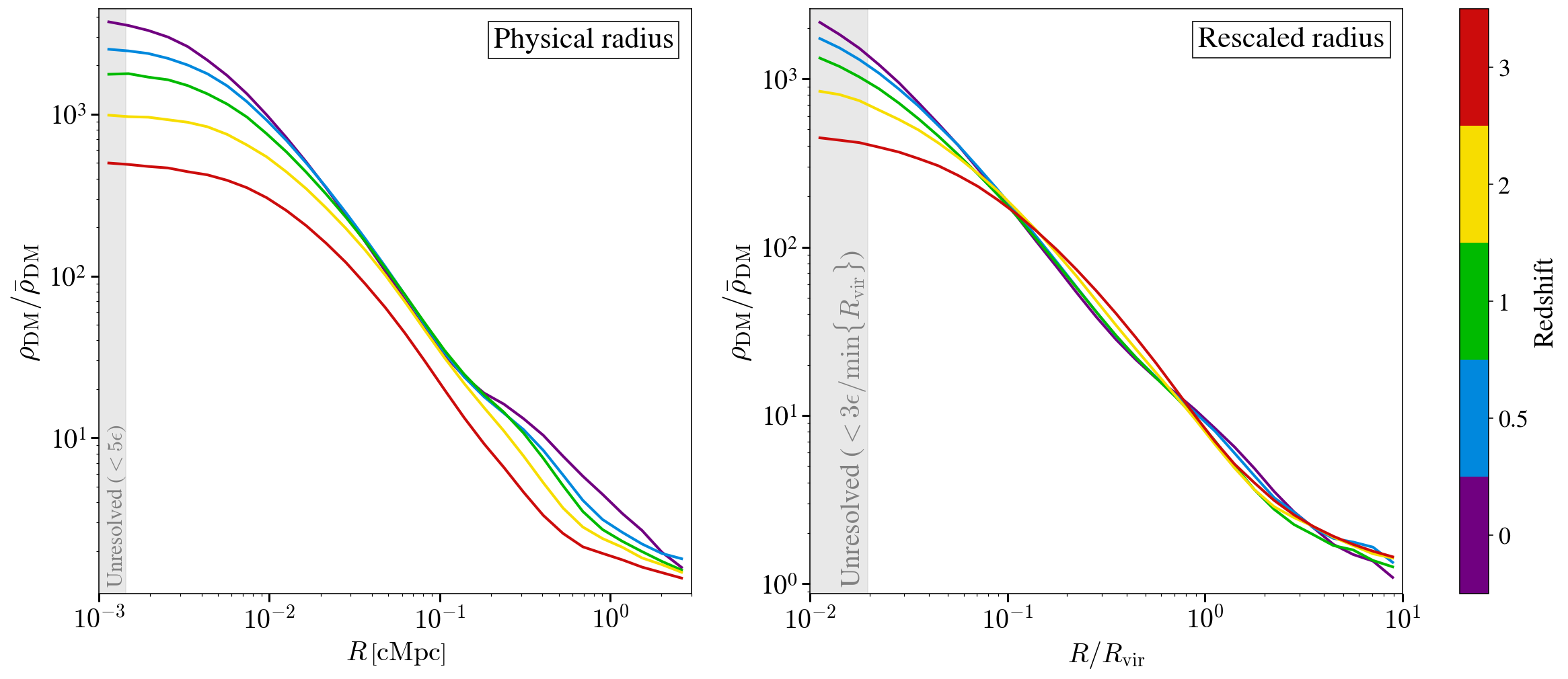}
    \end{subfigure}%
    \begin{subfigure}{0.295\textwidth}
        \centering
        \includegraphics[width=\linewidth]{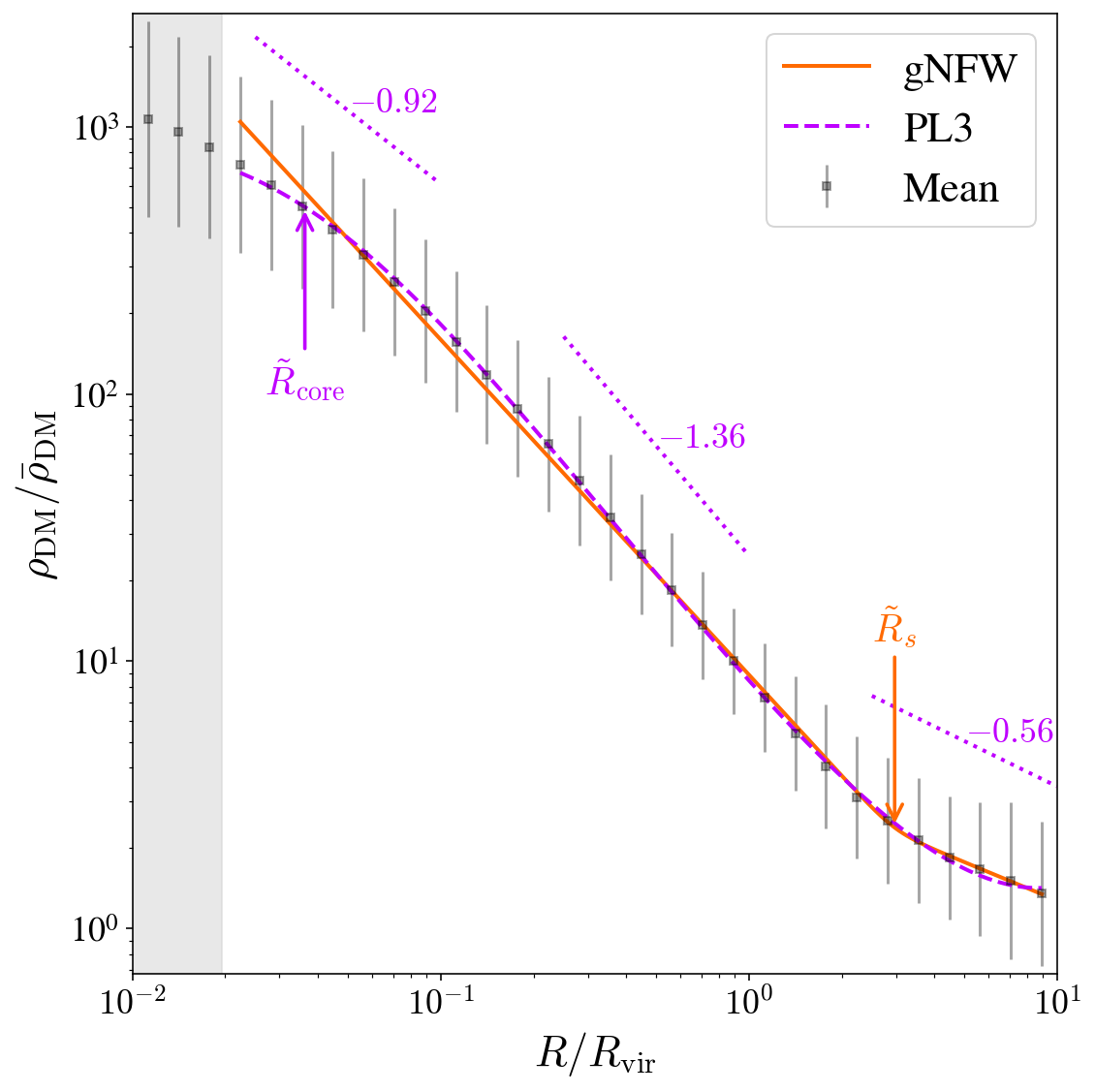}
    \end{subfigure}

    \caption{
    {\it Left}: DM density profiles of filaments in physical radius at different redshifts. 
    The unresolved region is shown in gray shades with the same definition as in \Fig{recentering}. 
    {\it Middle}: The same profiles, but with radii normalized by the virial radii ($\Rv$) of the connecting nodes. 
    Here, the gray area indicates a conservative global resolution limit, taken to be three times the DM softening length expressed in units of the virial radius of the lowest-mass node at $z = 3$.
    When normalized by the node virial radii, the mean profiles across redshifts become nearly universal down to $\sim 0.1 \Rv$.
    {\it Right}: Best fit to the filamentary density profile in rescaled radius.
    The profile is derived by stacking all filament samples across redshift and computing their length-weighted mean. 
    The error bars show the length-weighted standard deviation. 
    Solid and dashed curves correspond to the gNFW (Eq.(\ref{eq:gNFW})) and PL3 (Eq.(\ref{eq:PL3})) models, respectively. 
    The logarithmic density slopes in the core ($\tilde{R}\equiv R/\Rv\sim 0.05$), intermediate power-law regime ($\tilde{R}\sim 0.5$), and outskirts ($\tilde{R}\sim 5$) predicted by the PL3 model are shown as dotted guide lines.
    The gNFW scale radius $\tilde{R}_s$ and the PL3 core radius $\tilde{R}_{\mathrm{core}}$ are marked by arrows.
    Densities within the unresolved region (gray shade area) are excluded from the fitting procedure.
    }\label{fig:AllProfileAndBestFit}
\end{figure*}

The left and middle panels of \Fig{AllProfileAndBestFit} show the profiles across redshift and compare two representations: one with the radius in physical radius, the other with the radius normalized by the virial radius $\Rv$ of the connecting node.
Below, we refer to them as physical-radius profiles and rescaled profiles, respectively. 
Once the radius is normalized by the node virial radius, the redshift dependence of the average filament density profiles is largely removed. 
Overall, the profiles exhibit a nearly universal form down to 0.1$\Rv$. 

\subsection{Empirical models of the universal profiles}\label{sec:fit}

Despite the slight divergence in the central densities, we calculate a length-weighted mean density profile by stacking all the $\Rv$-rescaled filament profiles across all redshifts in order to characterize the universal profile shape. 
We first consider the generalized Navarro-Frenk-White model \citep[gNFW,][]{Zhao96, NFW97},
\begin{equation}\label{eq:gNFW}
    \frac{\rho}{\rho_0}=\frac{C_0}{\left(\frac{\tilde{R}}{\tilde{R}_s}\right)^\gamma\left[1+\left(\frac{\tilde{R}}{\tilde{R}_s}\right)^\alpha\right]^{\frac{\beta-\gamma}{\alpha}}},
\end{equation}
where $C_0$ is a normalization constant, $\tilde{R} \equiv R / R_{\mathrm{vir}}$ is the rescaled radius, and $\tilde{R}_s \equiv R_s / R_{\mathrm{vir}}$ is the corresponding dimensionless scale radius. 
The parameters $\gamma$ and $\beta$ represent the inner and outer logarithmic density slopes, respectively, while $\alpha$ controls the sharpness of the transition between them.

Because the filament profiles appear to exhibit three distinct regimes -- an innermost region ($\tilde{R} \la 0.1$) with a shallower logarithmic density slope than the immediately surrounding region, an intermediate power-law regime ($0.1\la  \tilde{R}\la 3$), and a flattened outskirts component -- we also consider a generalized triple-power-law model with two distinct scale radii \citep[PL3,][]{Yang23}, 
\begin{equation}\label{eq:PL3}
    \frac{\rho}{\rho_0}=\frac{C_{\mathrm{core}}}{1+(\frac{\tilde{R}}{\tilde{R}_{\mathrm{core}}})^\gamma(1+\frac{\tilde{R}}{\tilde{R}_{\mathrm{out}}})^\beta}.
\end{equation}
where $C_{\mathrm{core}}$ is a normalization constant, $\gamma$ controls the inner power-law slope, and $\beta$ determines the additional steepening at large radii, resulting in an asymptotic outer slope of $(\gamma+\beta)$, and $\tilde{R}_{\rm{core}}\equiv R_{\rm{core}}/R_{\rm{vir}}$ and $\tilde{R}_{\rm{out}}\equiv R_{\rm{out}}/R_{\rm{vir}}$ are the dimensionless inner and outer scale radii, respectively. 

\begin{deluxetable*}{lcc}
\tablecaption{Best-fit parameters of the gNFW (\eq{gNFW}) and PL3 ((\eq{PL3})) models for filament density profiles.\label{tab:BestFitModels}}
\tablehead{
\colhead{Parameter} &
\colhead{All DM particles} &
\colhead{Unbound DM particles}
}
\startdata
\multicolumn{3}{c}{\textbf{gNFW}} \\
\hline
$C_0$ 
& $2.27^{+0.07}_{-0.06}$
& $16.20^{+0.17}_{-0.20}$ \\
$\tilde{R}_s$ 
& $2.97^{+0.07}_{-0.08}$
& $0.33^{+0.00}_{-0.01}$ \\
$\alpha$ 
& $11.00^{+1.50}_{-1.25}$
& $-3.77^{+0.13}_{-0.14}$ \\
$\beta$ 
& $0.48^{+0.02}_{-0.02}$
& $0.09^{+0.01}_{-0.00}$ \\
$\gamma$ 
& $1.25^{+0.00}_{-0.01}$
& $1.02^{+0.01}_{-0.01}$ \\
$\chi^2_{\nu}$ 
& 0.016
& 0.006 \\
\hline
\multicolumn{3}{c}{\textbf{PL3}} \\
\hline
$C_{\mathrm{core}}$ 
& $993.98^{+31.97}_{-31.86}$
& $19.38^{+0.33}_{-0.26}$ \\
$\tilde{R}_{\mathrm{core}}$ 
& $0.04^{+0.00}_{-0.00}$
& $0.50^{+0.01}_{-0.01}$ \\
$\tilde{R}_{\mathrm{out}}$ 
& $39.08^{+14.64}_{-9.93}$
& $42.63^{+101.01}_{-22.52}$ \\
$\beta$ 
& $-8.15^{+1.75}_{-2.78}$
& $-11.97^{+4.92}_{-30.50}$ \\
$\gamma$ 
& $1.50^{+0.01}_{-0.01}$
& $1.79^{+0.03}_{-0.45}$ \\
$\chi^2_{\nu}$ 
& 0.002
& 0.001 \\
\enddata
\tablecomments{
The PL3 model is designed to accurately fit only the inner part of the profiles. As a result, the best-fit solution may formally produce a negative outer slope, $(\gamma + \beta)$, at an extremely large outer scale radius, $\tilde{R}_\mathrm{out}$.
Such outer-profile parameters lie outside the intended fitting regime and should not be assigned physical significance.
}
\end{deluxetable*}

To estimate the uncertainties of the best-fit model parameters, we apply a bootstrap resampling procedure. 
For each bootstrap realization, we randomly draw the same number of filaments from the catalog, compute the corresponding length-weighted mean profile, and fit either the gNFW or PL3 model to the profile.
This is repeated 500 times and the best-fit parameters for all realizations are recorded.
We quote the $1\sigma$ uncertainty of each parameter using the 16th and 84th percentiles of the bootstrap distribution.
The best-fit parameters for both models are listed in the left column of \Tab{BestFitModels}.  
The data points and the models using the best-fit parameters of the gNFW and PL3 model are displayed in the right panel of \Fig{AllProfileAndBestFit}. 
The PL3 model gives a negative outer slope and a outer scale radius much larger than the scales of interest here, which indicates that the outer slope and radius of the PL3 model are not relevant. 
This will also be seen in the fit to the profile constructed from unbound DM particles (\se{unbound}).
According to the best-fit PL3 model, the logarithmic density slopes in the core ($\tilde{R}\sim 0.05$), intermediate power-law regime ($\tilde{R}\sim 0.5$), and outskirts ($\tilde{R}\sim 5$) are -0.92, -1.36, and -0.56, respectively.
Since both models have the same number of degrees of freedom, the substantially smaller $\chi^2_{\nu}$ for the PL3 model demonstrates that it is s superior empirical model that more accurately describes filament density profiles.

\section{Discussion}\label{sec:discussion}

\subsection{Dependence of filament profiles on node mass, length, and redshift}
\label{sec:dependences}

\begin{figure*}
    \centering

    \begin{subfigure}{\textwidth}
        \centering
        \includegraphics[width=\textwidth]{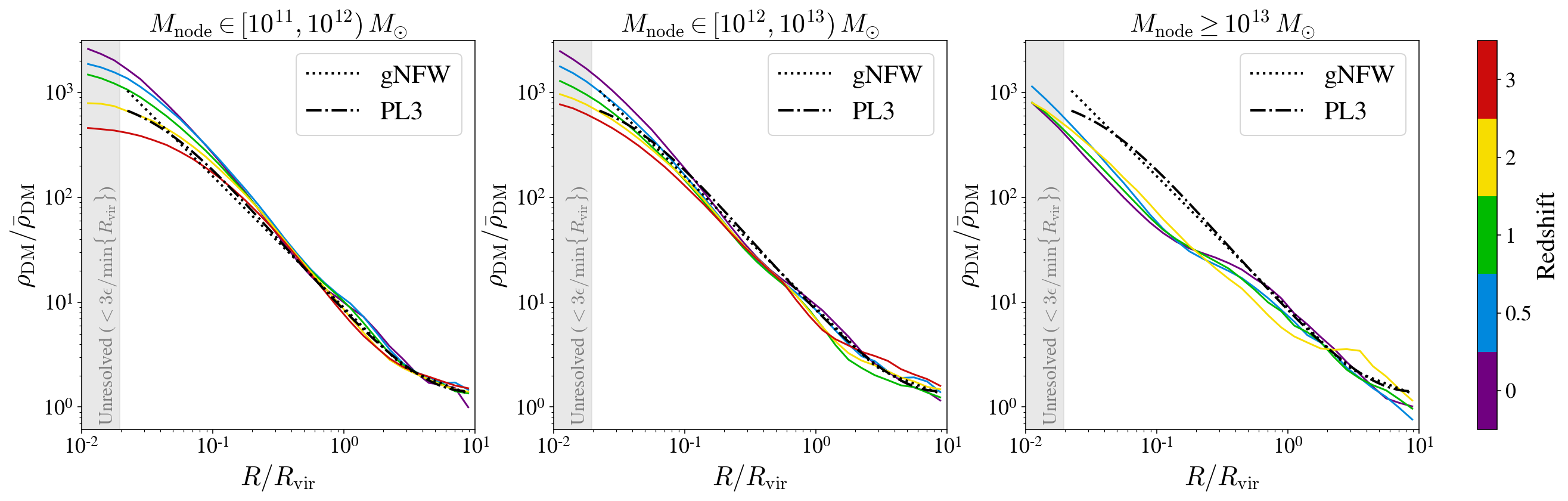}
    \end{subfigure}

    \vspace{0.5em}

    \begin{subfigure}{\textwidth}
        \centering
        \includegraphics[width=\textwidth]{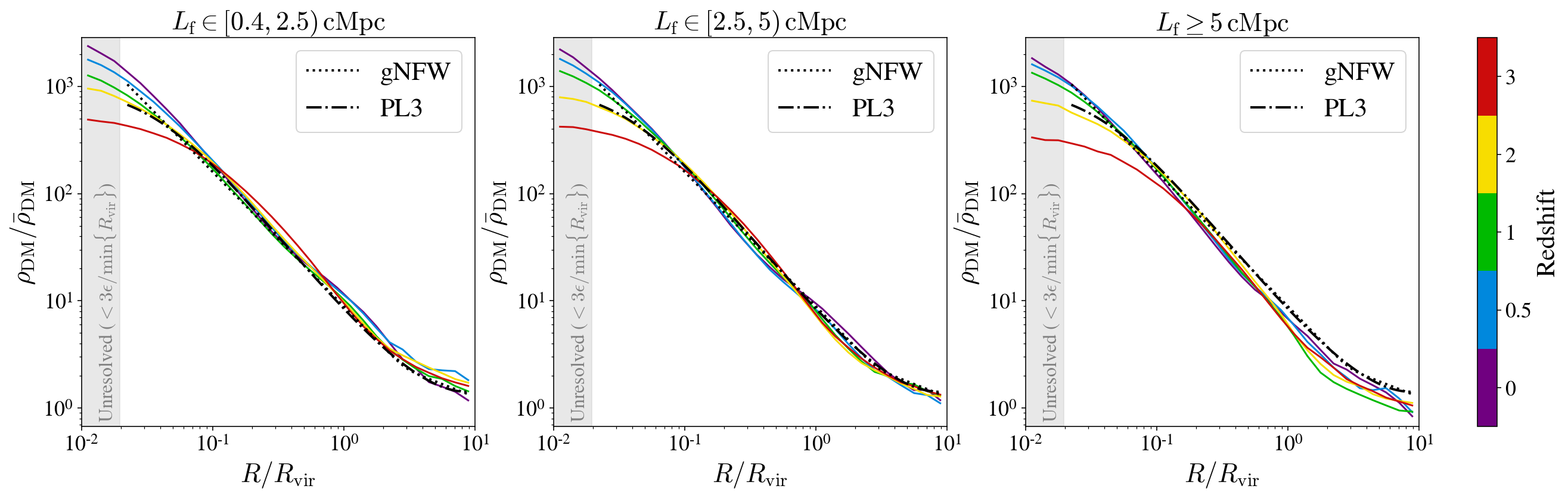}
    \end{subfigure}

    \caption{Dark-matter density profiles around filaments as functions of node mass (top) and filament length (bottom). 
    In each panel, the length-weighted mean profiles across redshifts are shown for the mass bin or length bin indicated. 
    The best-fit gNFW (dotted) and PL3 (dashed) models for the full sample, as in \Fig{AllProfileAndBestFit} and \Tab{BestFitModels}, are overlaid for reference. 
    Note that no filaments are present in the $M_{\mathrm{node}} > 10^{13} M_{\odot}$ bin at $z=3$.
    \quad
    Overall, the node-mass or length dependence is weak, except in the highest-mass bin where the result is affected by small-number statistics. 
    There is a weak redshift dependence in the central density at $R/\Rv \la 0.1$, with higher-redshift filaments exhibiting lower central densities.}
    \label{fig:MassAndLengthDependence}
\end{figure*}

With the overall DM density profile accurately modeled, we proceed to examine its dependence on the mass of the connecting nodes, filament length, and redshift.
The top row of \Fig{MassAndLengthDependence} presents the length-weighted mean filament DM density profiles in three node-mass bins.  
Overall, the profiles show only a weak dependence on node mass, except that  filaments associated with nodes more massive than $10^{13}\Msun$ exhibit a suppression in the inner density profile at $R/\Rv \la 1$.  
We caution against over-interpreting this feature, as the high-mass bin contains only a small number of filaments, particularly at higher redshifts (see the left panel of \Fig{NodeMassAndLengthDistribution} in Appendix \ref{AppendixA} for the node mass distribution). 

At a fixed mass, there is also a weak redshift dependence of the central DM density, in the sense that the density is lower at higher redshift. 
This effect is relatively more pronounced in the lowest node-mass bin, $10^{11-12}\Msun$, where the central density difference between $z=0$ and $3$ reaches up to 0.5 dex. As will be discussed in \se{unbound}, this difference can be attributed to the difference in the amount of small-scale halos near filament spine across redshifts.

To probe the dependence on filament length, we classify the filaments into three categories: short ($L_{\rm{f}}<2.5 \,\rm{cMpc}$), intermediate ($2.5\,\mathrm{cMpc} \leq L_\mathrm{f} < 5\,\mathrm{cMpc})$, and long ($L_{\rm{f}} \geq 5\, \rm{cMpc})$. 
These boundaries are motivated by the filament length distributions shown in the right panel of \Fig{NodeMassAndLengthDistribution} in Appendix \ref{AppendixA}.
The bottom row of \Fig{MassAndLengthDependence} shows the length-weighted mean filament density profiles in the three length bins and reveals a negligible dependence on filament length. 

\subsection{The halo contribution to inner filament density profiles}\label{sec:unbound}

The hierarchical nature of structure formation implies that smaller, earlier-forming halos are embedded within cosmic filaments that feed the growth of later generations of larger nodes. 
It is therefore natural to hypothesize that the high central density of the filaments of $ \sim 10^3 \bar{\rho}_{DM}$ are contributed by these halos along filament spines. 
To quantify the halo contribution to the filament density profiles, we remove all DM particles bound to resolved central halos with masses above $10^8 \Msun$ along the filaments.

Removing these particles from the density measurement involves subtracting both their mass and the volume they occupy. The latter is non-trivial because halos can spatially overlap with one another.
We therefore adopt the following procedure. 
First, the central halos are selected using the FoF groups' {\tt GroupFirstSub} field in the \textsc{SUBFIND} catalog. 
Second, we perform a Monte-Carlo sampling within the cylindrical shells surrounding each filament segment to estimate the volume occupied by halos. 
Specifically, random points are uniformly sampled within each shell, and the fraction of points falling within the virial radius of any central halo is used to estimate the halo-occupied volume, which is then subtracted from the shell volume.
Finally, we remove the mass contribution of bound particles using their halo membership information provided by the TNG simulation.
Through the process, both the mass and occupied volume of corresponding halos are removed consistently. 

\begin{figure*}
    \centering

        \includegraphics[width=0.4\linewidth]{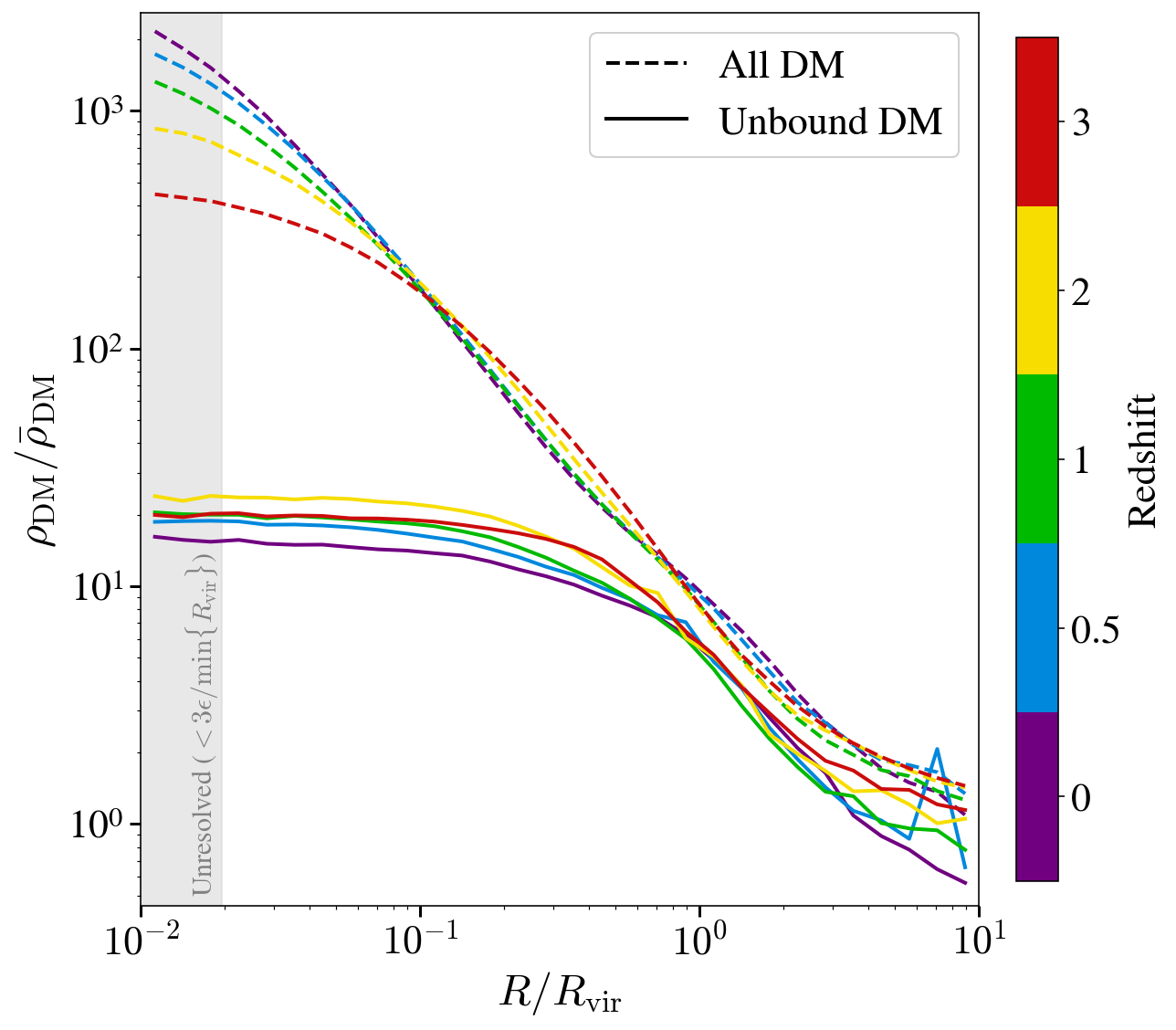}
        \includegraphics[width=0.35\linewidth]{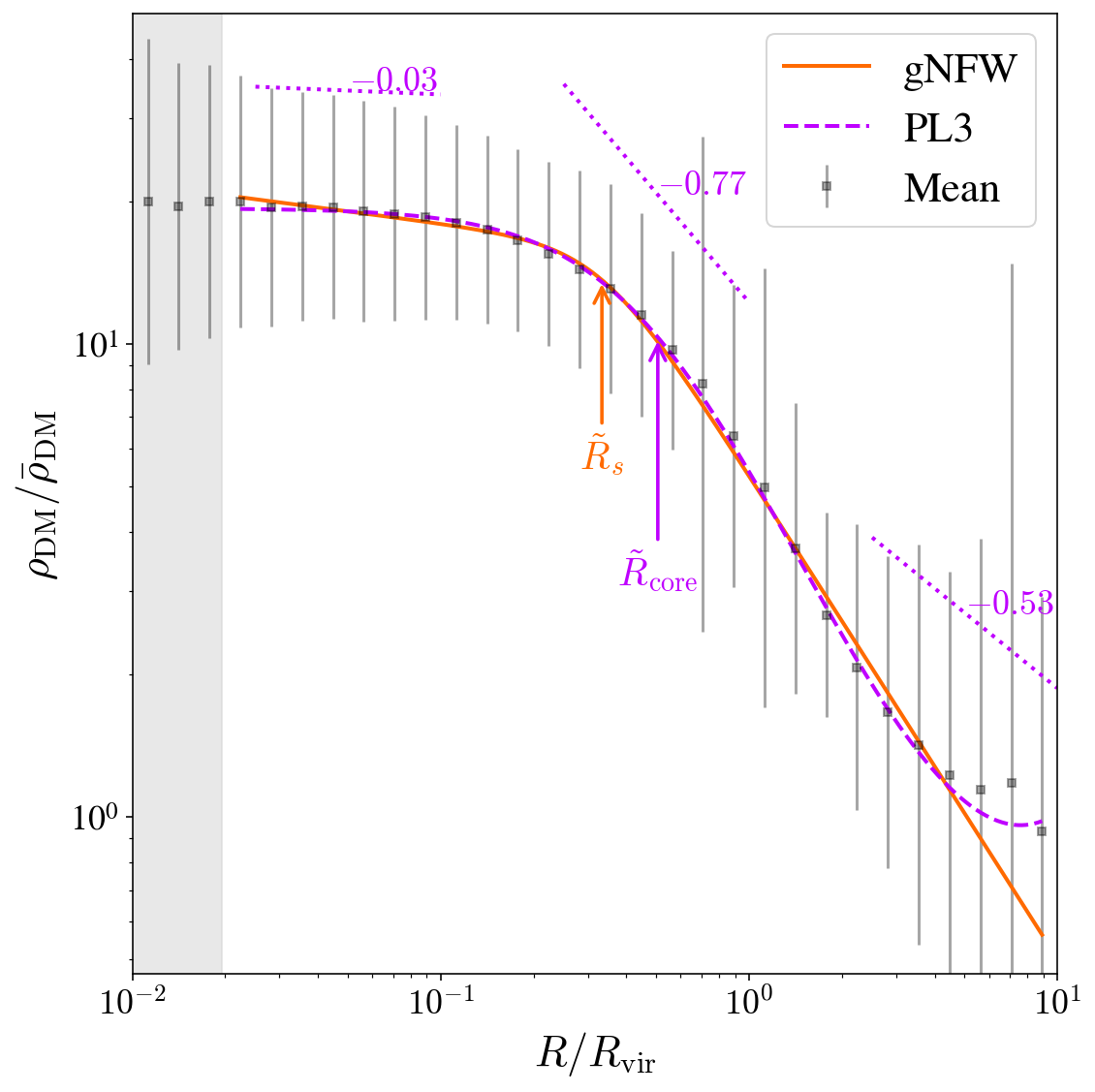}

    \caption{
    Impact of halos near the filament spine on the DM density profiles. 
    \textit{Left}: Comparison of length-weighted mean filament density profiles calculated using all DM particles (dashed) and unbound DM particles (solid). 
    Here the ``unbound DM'' profile is constructed by removing all DM particles bound to central halos with mass above $10^8M_{\odot}$ together with the volumes occupied by these halos.
    The resulting smooth filaments exhibit flat density cores whose central densities are reduced by up to 2 dex relative to the original clumpy filaments, indicating that low-mass halos dominate the central regions within $R/R_{\mathrm{vir}}\sim 0.1$.
    \quad
    \textit{Right}: Length-weighted mean filament density profiles stacked over all five epochs considered here, traced by unbound DM.
    Solid and dashed curves correspond to the best-fit gNFW (Eq.(\ref{eq:gNFW})) and PL3 (Eq.(\ref{eq:PL3})) models, respectively. 
    The logarithmic density slopes in the core ($\tilde{R}\equiv R/\Rv\sim 0.05$),  intermediate power-law regime ($\tilde{R}\sim 0.5$), and outskirts ($\tilde{R}\sim 5$) are labeled and indicated by dotted guide lines.
    The gNFW scale radius $\tilde{R}_s$ and the PL3 core radius $\tilde{R}_{\mathrm{core}}$ are marked by arrows.
    Densities within the unresolved region (gray shaded area) are excluded from the fitting procedure.
    }
    \label{fig:UnboundProfileAndBestFit}
\end{figure*}

As shown in the left panel of \Fig{UnboundProfileAndBestFit}, the density profiles constructed from smoothly distributed, unbound DM particles are significantly reduced relative to the original all-particle profiles. 
This difference is especially pronounced in the central regions, reaching up to 2 dex, where the smooth component exhibits a flat core at $R/\Rv\la 0.1$.
Since $\Rv\propto\Mv^{1/3}$ at each epoch, this implies that the regions closest to the filament spines are predominantly contributed by halos with masses  $\la10^{-3}$ of the node mass.

The redshift trend of the central density also reverses sign: the smooth DM density now decreases towards lower redshift. 
This suggests that, at higher redshift, cosmic nodes accrete from filaments more strongly through smooth accretion, whereas at later times the feeding filaments become increasingly clumpy, containing a larger mass fraction in halos.  
This interpretation is consistent with previous studies of halo merger trees that explicitly track a smooth-accretion component in addition to halo mergers \citep{Cole00,Parkinson08,Jiang14}, all of which find that the smooth-accretion fraction increases towards higher redshift. 

We also repeat the same profile fitting for the smooth filaments. 
The best-fit parameters for the gNFW and PL3 models are listed in the right column of \Tab{BestFitModels}, and the corresponding fits are shown in the right panel of \Fig{UnboundProfileAndBestFit}.
According to the best-fit PL3 model, the logarithmic density slopes in the core ($\tilde{R}\sim 0.05$), intermediate power-law regime ($\tilde{R}\sim 0.5$), and outskirts ($\tilde{R}\sim 5$) are -0.03, -0.77, and -0.53, respectively.
At radii far from the spine, the profile is less affected by halo contamination, and the two outer slopes therefore remain close to those of the all-particle case.

\subsection{Comparison with galaxy-traced filament profiles}

\begin{figure*}
    \includegraphics[width=\textwidth]{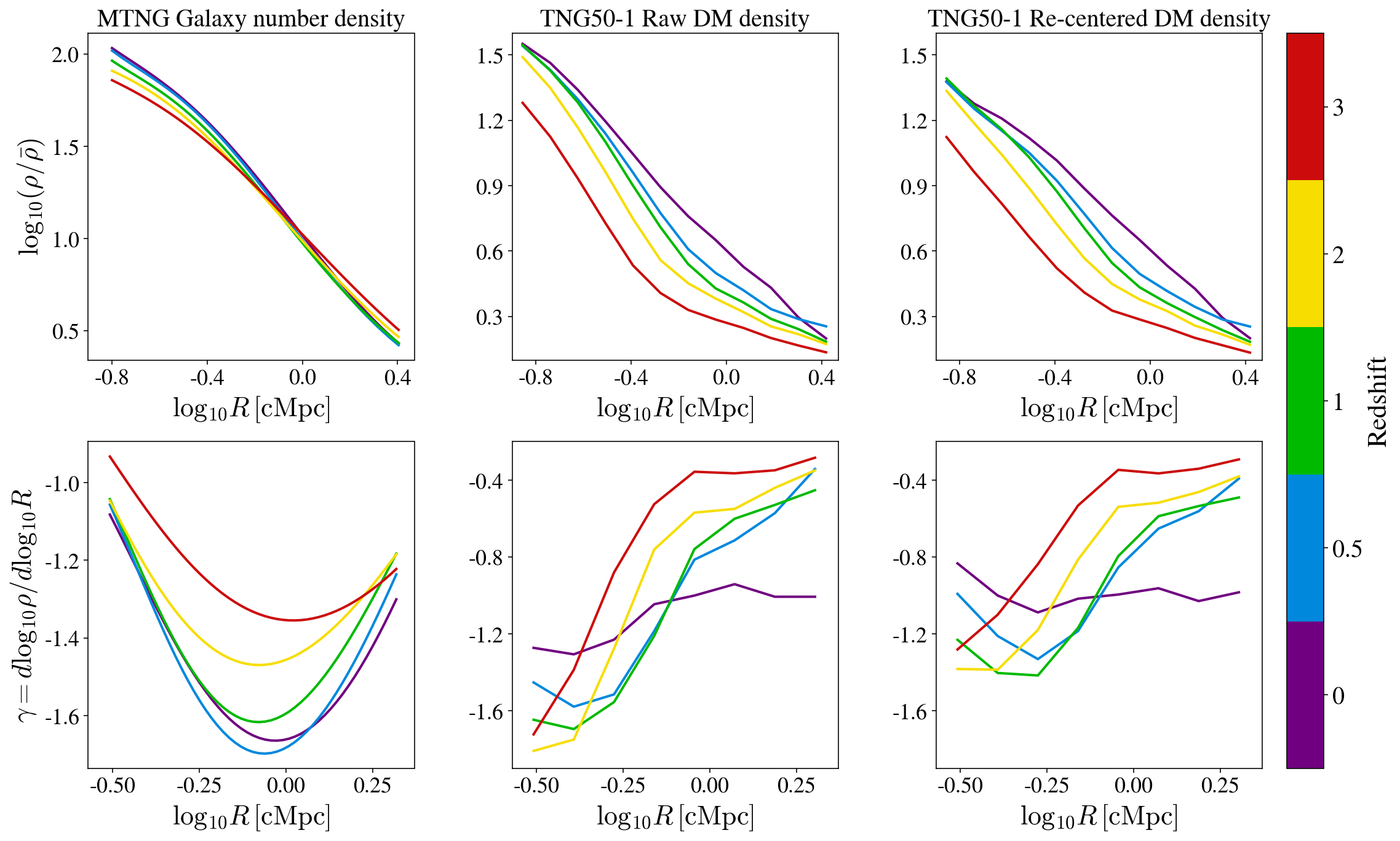}
    \caption{
    Comparison of length-weighted mean filament profiles from this work with those reported by \citet{Wang24}, which used galaxies in the MTNG simulation to identify and characterize filaments. 
    The upper and lower rows present the density profiles and the logarithmic density slopes, respectively. 
    The left column shows the \citeauthor{Wang24} results. 
    The middle column shows our results measured around the raw filament spines, i.e., without applying the shrinking cylinder re-centering. 
    The right column shows the corresponding profiles using the re-centered filaments.
    \quad
    Over the overlapping radial range $R \simeq$ 0.16–2.5 cMpc, the MTNG galaxy-traced filaments are denser by up to $\sim$ 0.5 dex and exhibit a non-monotonic slope profile with a characteristic scale near 1 cMpc, whereas our DM-based profiles flatten monotonically with a smaller characteristic scale of $\sim$ 0.6 cMpc, where the slope flattens toward larger radii.
    }
    \label{fig:Wang24Comparison}
\end{figure*}

Density profiles and characteristic radii of cosmic filaments have previously been reported by \cite{Wang24} using the MillenniumTNG simulation \citep[MTNG,][]{Hernandez-Aguayo23, Kannan23, Pakmor23}.
Here, we compare our filament density profiles with their results.
We emphasize, however, that a fundamental methodological difference exists between \citeauthor{Wang24} and the present work. Their filaments are identified using galaxies (with stellar masses greater than $10^9M_{\odot}$), and the corresponding profiles are measured from galaxy number density, whereas in this work we directly use the DM density field.
Because galaxies are biased tracers of the underlying matter density field, exact agreement is not expected. 
The comparison therefore primarily serves to gauge the impact of using biased tracers. 

The results are presented in \Fig{Wang24Comparison}: the upper and lower panels show the density profiles and the logarithmic density slopes, respectively, both as functions of the physical distance to the filament spine over the overlapping radial range $R\simeq 0.16$ - 2.5 cMpc. 
Substantial differences are evident. 
First, the galaxy-traced filaments are systematically denser by up to $\sim 0.5$ dex.
Second, the profile shapes differ significantly. 
The galaxy-traced filaments exhibit a non-monotonic density slope profile, steepening from $-1$ to $\sim -1.5$ at a characteristic scale of $\sim 1$ cMpc before flattening again at larger radii. 
By contrast, our DM-based filaments show a nearly monotonic flattening of the density slope toward the outskirts, except in the $z=0$ cases, which remain consistent with a constant power-law slope of $\sim -1$.
If one defines an analogous characteristic scale in our profiles based on the change in slope, the radius is $\sim 0.6$ cMpc, where the profile begins to flatten toward larger radii. 

The differences are not surprising given the distinct tracers used to construct the profiles. 
Equally important, however, is the fact that the filaments probed in these two studies span substantially different physical scales.
The MTNG analysis extracts only the largest-scale filaments. 
This is partly because \citeauthor{Wang24} apply DTFE to the point set consisting of galaxies, and adopt a fixed persistence threshold of ${\tt nsig}=2$ across all redshifts, and partly because MTNG itself has a large box size but comparatively low spatial resolution. 
In contrast, by exploiting a finely gridded density field in a higher-resolution simulation, we trace filaments down to much smaller scales. 

In addition to \citeauthor{Wang24}, several other studies have also investigated the DM and/or gas density profiles of cosmic filaments \citep{Galarraga-Espinosa20, Galarraga-Espinosa21, Zhu21, Galarraga-Espinosa24, Bahe25, Yang25}. A more systematic and comprehensive comparison with these studies is beyond the scope of the present paper, but represents an important direction for our future work.

\section{Conclusions}\label{sec:conclusion}

In this paper, we have presented a comprehensive analysis of the dark-matter density profiles of cosmic filaments at $z = 0$, 0.5, 1, 2 and 3 in the cosmological simulation TNG50-1. Using the DisPerSE algorithm on high-resolution density grids, together with a new shrinking-cylinder spine re-centering algorithm, we establish a robust pipeline for accurately measuring filament radial profiles across cosmic time.
Our main findings are as follows.

\begin{itemize}[leftmargin=*]

\item 
The filament spines returned directly by DisPerSE are offset from the true density ridges because of the finite resolution of the input density grid.  
This leads to artificial flattened cores in the filament profile. 
We develop a shrinking-cylinder algorithm to refine the spine locations.
Correcting this bias with our shrinking-cylinder algorithm increases the inferred central densities by up to 1–1.5 dex and restores the inner power-law behavior of the profiles.
Notably, the central densities of the raw profiles agree well with recent studies \citep[e.g.][]{Bahe25}, indicating that this artifact has remained uncorrected for a long time. 

\item 
We find that filament density profiles exhibit near universality when scaled by the node virial radius. 
When expressed as a function of $\tilde{R}=R/\Rv$, the redshift dependence of the mean profiles is largely removed, revealing a nearly universal filament density profile across $z=0$–3. 
Previous studies have attempted to establish a similar self-similarity by introducing the concept of filament width, defined as the radius at which the radial density profile declines to the cosmic mean. 
Profiles normalized by this width show a weak degree of universality in the outer regions, when examined within bins of width and redshift \citep{Zhu21, Yang25}. 
In contrast, the scaling adopted here leads to a significantly stronger universality over a broader radial range, suggesting that the node virial radius provides a more physically meaningful characteristic scale for filamentary structures.
This universality depends only weakly on node mass and filament length.
Hence, cosmic filaments, much like dark-matter halos, obey a form of structural self-similarity once an appropriate characteristic scale is introduced. 

\item 
We find that a generalized triple-power-law model provides an accurate empirical description of the universal density profiles.
On average, a filament has three characteristic regimes: an inner flattened region within $\tilde{R}\sim 0.2$, an intermediate power-law segment at $0.2 \la \tilde{R}\la 3$, and a shallower outskirts component at $\tilde{R}\ga 3$. 

\item 
We notice that the innermost regions of filaments are dominated by low-mass halos.
After removing halo-bound particles and subtracting the corresponding occupied volumes, the filament profiles show a flat core within $\tilde{R}\la 0.1$, with central densities reduced by up to $\sim 2$ dex. 
This indicates that the apparent central cusp in the all-particle profile is primarily contributed by halos with masses $\la 10^{-3}$ of the node mass.
The smooth filaments exhibit higher central densities toward higher redshift, implying that high-redshift node growth is fed more by smooth filamentary accretion, while later filaments become increasingly clumpy as a larger fraction of their mass collapses into halos prior to node infall.

\end{itemize}

Despite these advances, several challenges in this field remain unsolved. 
In this work, we use the smallest yet highest-resolution box in the IllustrisTNG suite to capture filamentary structures across multiple scales and to characterize their structures in detail. 
However, the large-scale nature of the cosmic web demands simulations with larger volumes, while still maintaining high resolution, in order to obtain a more complete filament sample. 
In addition, the widely used DisPerSE algorithm relies on a set of somewhat ad hoc, non-physical parameters for filament extraction, and there is no strict or transparent mapping between these parameters and the physical scales of interest. 

Building on this work, we plan to continue to pursue a more physical characterization of cosmic filaments. 
As noted in \se{intro}, density estimates obtained with DTFE do not reliably recover the underlying true densities. 
Yet, any attempt to reconstruct the real cosmic web must ultimately rely on galaxy tracers.
Recent studies have made substantial progress in this direction, including the improved Monte Carlo Physarum Machine \citep[MCPM,][]{Hasan24} and machine-learning–based approaches \citep{Wang23, Shi25}. 
A detailed comparison between MCPM-derived filament profiles and those measured directly from simulations will be presented in a forthcoming work. 
With increasingly robust filament extraction and modeling, our longer-term goal is to establish how cosmic filaments regulate galaxy formation and evolution.

\begin{acknowledgments}
We dedicate this work to the memory of our co-author Joel R. Primack, whose contributions played key roles in forming the foundation of the theory of cold dark matter (CDM) and galaxy formation. 
We acknowledge the assistance on software manuals of Xuelin Chen from the School of Computer and Data Science at HKU. 
We also acknowledge insightful and interesting discussions with Sandra M. Faber, Peng Wang, Wei Wang, Dandan Xu, Daisuke Nagai and Song Huang.
FJ acknowledges support by the National Natural Science Foundation of China (NSFC, 12473007) and China Manned Space Program with grant no. CMS-CSST-2025-A03.
\end{acknowledgments}

%



\appendix

\setcounter{figure}{0}
\renewcommand{\thefigure}{A\arabic{figure}}
\setcounter{table}{0}
\renewcommand{\thetable}{A\arabic{table}}
\renewcommand{\theHtable}{A\arabic{table}}

\section{Details of DisPerSE filaments}
\label{AppendixA}

\begin{figure}
    \centering
    \begin{subfigure}{0.37\textwidth}
        \centering
        \includegraphics[width=\linewidth]{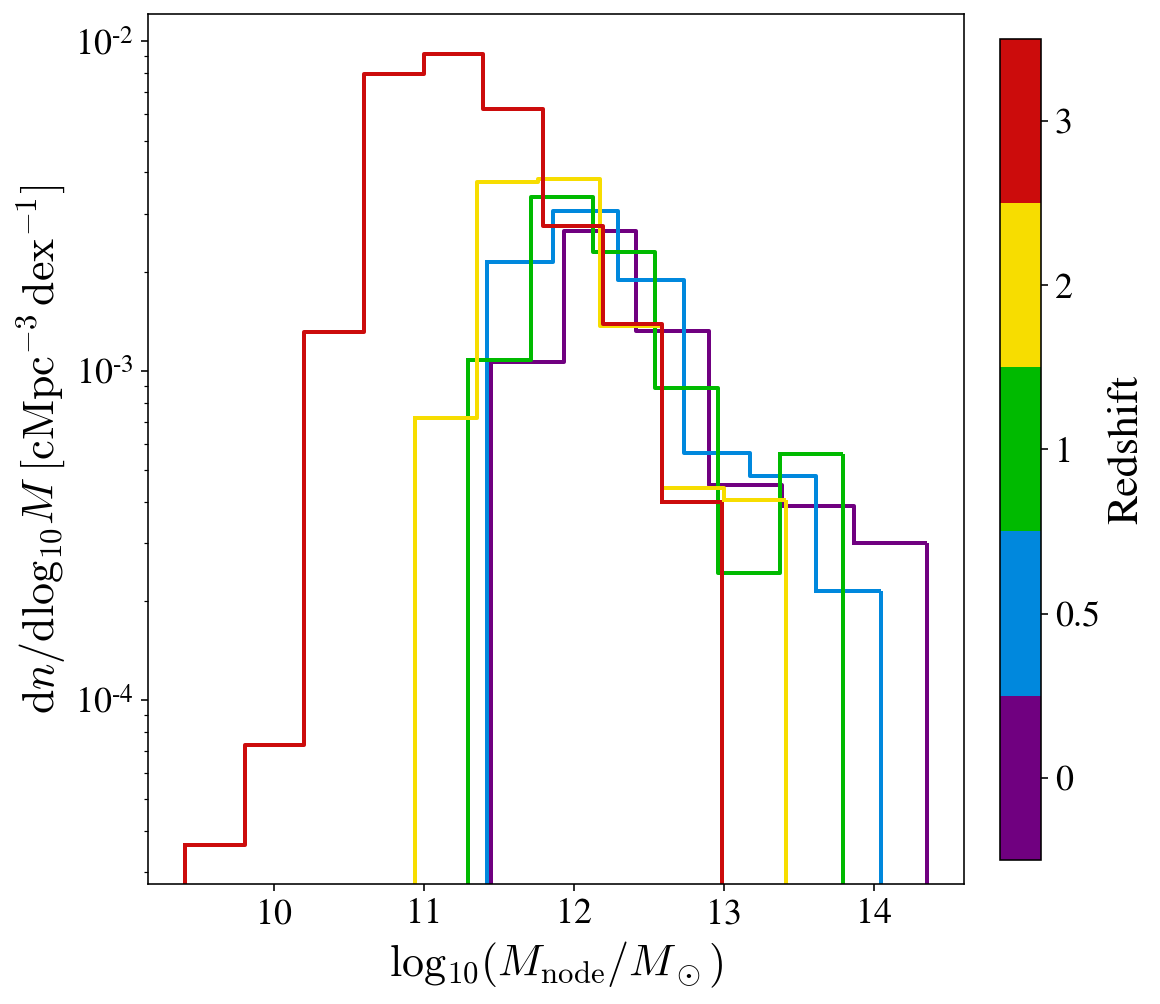}
    \end{subfigure}%
    \begin{subfigure}{0.37\textwidth}
        \centering
        \includegraphics[width=\linewidth]{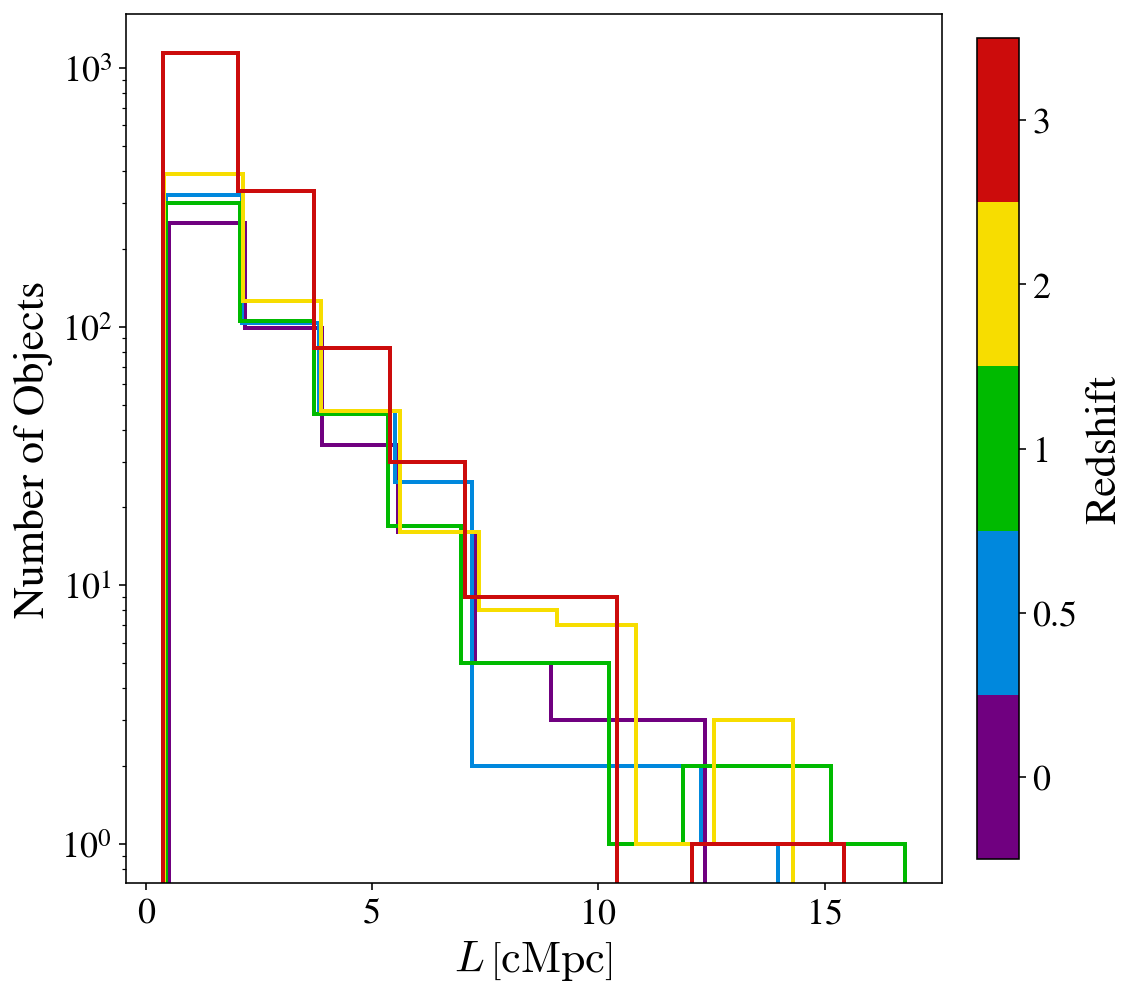}
    \end{subfigure}

    \caption{
    {\textit{Left}}: Mass function of the FoF groups associated with nodes at different redshifts.
    {\textit{Right}}: Filament length distribution at different redshifts. By definition, filaments are sets of contiguous segments connecting maximum to 2-saddles (see Section \ref{sec:DisPerSE}).
    }
    \label{fig:NodeMassAndLengthDistribution}
\end{figure}

The mass function of nodes associated with CPmax and the distribution of filament lengths are displayed in \Fig{NodeMassAndLengthDistribution}. The parameter choices of DisPerSE and the basic statistics of filaments are shown in \Tab{skeleton_details}.

\begin{deluxetable}{lccccc}
\tablecaption{Details of skeletons at different redshifts\label{tab:skeleton_details}}
\tablehead{
\colhead{Redshift} &
\colhead{0} &
\colhead{0.5} &
\colhead{1} &
\colhead{2} &
\colhead{3}
}
\startdata
{\tt{cut}} (Persistence)        & 15  & 13  & 12  & 5   & 1    \\
Number of filaments      & 469 & 595 & 585 & 769 & 1965 \\
Number of CPmax          & 243 & 309 & 309 & 413 & 1201 \\
Mean length [cMpc]       & 2.3 & 2.1 & 2.1 & 1.9 & 1.6  \\
Max length [cMpc]        & 10.8& 12.1& 15.3& 12.5& 13.4 \\
Min length [cMpc]        & 0.4 & 0.4 & 0.5 & 0.4 & 0.4  \\
\enddata
\end{deluxetable}

\section{Comparison of the density fields by all and unbound DM particles} \label{Appendix B}

\begin{figure*}
    \centering
    \includegraphics[width=0.9\textwidth]{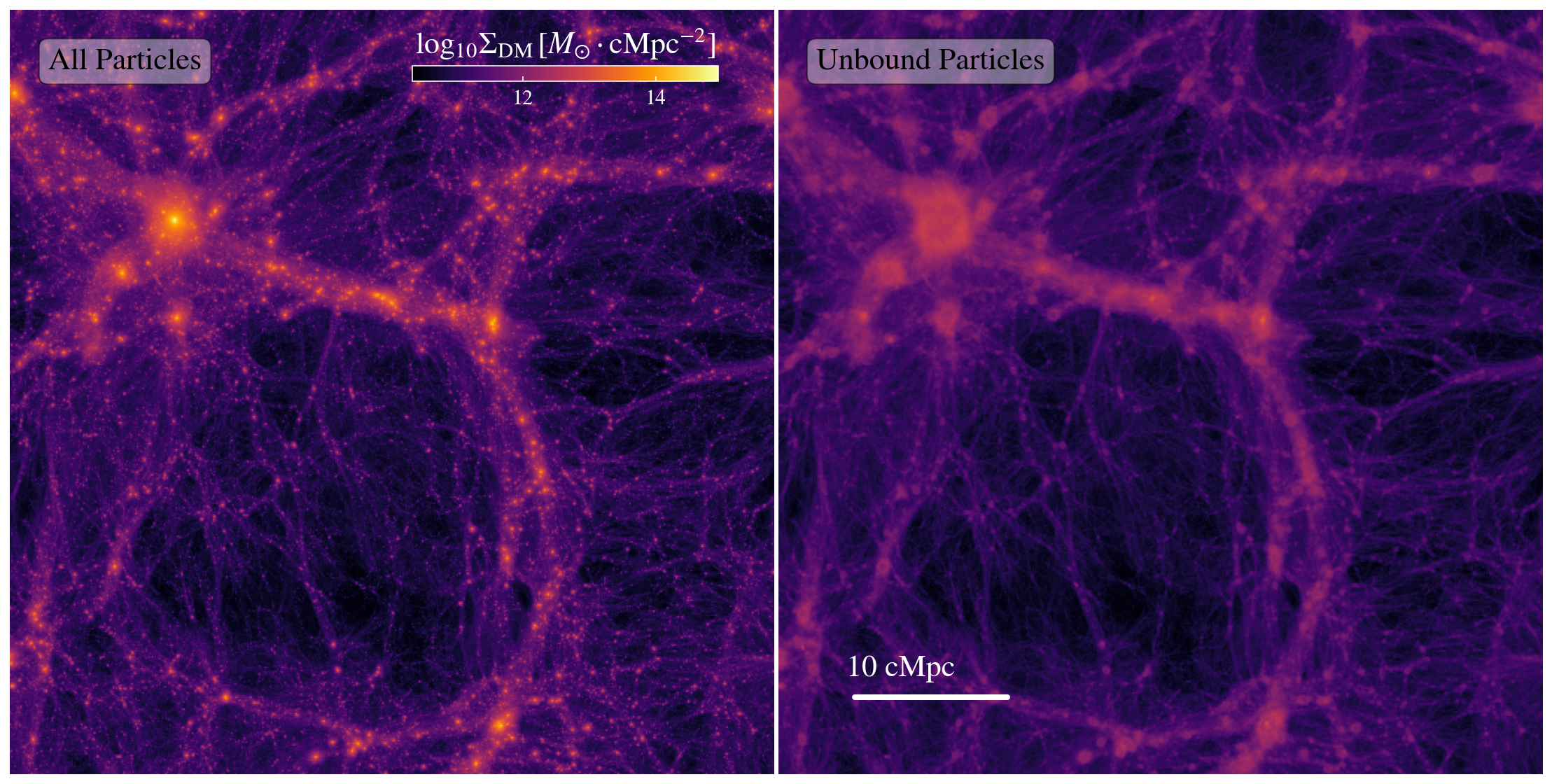}
    \caption{Comparison between the distribution of all DM particles and unbound DM particles at $z$ = 0 in TNG50-1. 
    {\textit{Left}}: The density projection of all DM particles. The full 3D box has been projected onto the x-y plane. 
    {\textit{Right}}: The same projection of DM particles unbound to any subfind halo.}
    \label{fig:UnboundDMDensity}
\end{figure*}

\Fig{UnboundDMDensity} explicitly demonstrates the effect of removing halo-bound DM particles.

\bibliography{draft}{}
\bibliographystyle{aasjournalv7}



\end{document}